\pdfoutput=1
\UseRawInputEncoding
\documentclass[11pt]{article}

\usepackage[preprint]{acl}

\usepackage{times}
\usepackage{latexsym}

\usepackage{pifont}

\usepackage{graphicx} % Required for inserting images
\usepackage[frozencache,cachedir=.]{minted}
\usepackage{xcolor}
\usepackage{url}
\usepackage{subcaption}
\usepackage{amsmath}
\usepackage{cuted}

\usepackage[T1]{fontenc}    % use 8-bit T1 fonts
\usepackage{hyperref}       % hyperlinks
\usepackage{amsfonts}       % blackboard math symbols
\usepackage{nicefrac}       % compact symbols for 1/2, etc.
\usepackage{microtype}      % microtypography
\usepackage{natbib}
\usepackage{doi}

\definecolor{keycolor}{HTML}{058202}
\usepackage{tcolorbox}

\usepackage{tabularx}

\usepackage{booktabs}
\graphicspath{{figs/}}

\newcommand{\nop}[1]{}

\title{ADL: A Declarative Language for Agent-Based Chatbots}
\author{Sirui Zeng \\
  \texttt{sirui\_zeng@ucsb.edu} \\\And
  Xifeng Yan \\
  \texttt{xyan@cs.ucsb.edu} \\}
\begin{document}
\maketitle
\begin{abstract}
There are numerous frameworks capable of creating and orchestrating agents to address complex tasks. However, most of them highly coupled Python programming with agent declaration, making it hard for maintenance and runtime optimization.  In this work, we introduce ADL, an agent declarative language for  customer service chatbots. ADL abstracts away implementation details, offering a declarative way to define agents and their interactions,  which could ease maintenance and debugging. It also incorporates natural language programming at its core to simplify the specification and communication of chatbot designs. ADL includes four basic types of agents and supports integration with custom functions, tool use, and third-party agents. MICA, a multi-agent system designed to interpret and execute ADL programs, has been developed and is now available as an open-source project at \url{https://github.com/Mica-labs/MICA}. Its documentation can be found at \url{https://mica-labs.github.io/}. 
\end{abstract}

\section{Introduction}
Task-oriented dialogue (TOD) systems can assist users in completing tasks such as flight change or online shopping. With the advancement of large language models (LLMs), the development of ToD systems has become significantly simpler, leading to substantial improvements in both efficiency and performance. Components such as NLU, Dialogue State Management, Dialogue Policy, and NLG \citep{chen2017survey}, which previously required separate implementations and fine-tuning, can now be effectively integrated and executed through LLMs \citep{madotto2020language, bae2022building}. In some cases, the entire dialogue workflow can be encapsulated within a single prompt, simplifying the design process while maintaining robust conversational capabilities.

While it is possible to pack all constraints, business logic, and knowledge into a single LLM prompt (essentially creating one gigantic agent), this approach can lead to issues in development, debugging, maintenance, and reusability as  business logic evolves over time. \nop{For example, air ticket booking could be a quite standard routine for many airlines. However, the membership promotion and mileage redemption program of different airlines could be very different and need to be customized individually.}Hence, it is critical to have a multi-agent design for customer service bot development. 

There are numerous agent frameworks available, such as AutoGen~\citep{autogen2023}, LangGraph~\citep{LangGraph}, CrewAI~\citep{crewAI}, Swarm~\citep{Swarm}(OpenAI Agents SDK~\citep{agentsdk}), and Multi-Agent Orchestrator~\citep{MAO} which provide high flexibility for creating and orchestrating agents in general settings. However, they seem overly complicated for customer service chatbots.  Their agent declarations are often interleaved within intricate Python code, making it difficult to grasp the entire design and hard to do runtime optimization. 

% \begin{figure*}[t!]
%     \centering
%     \begin{subfigure}[b]{0.4\textwidth}
%         \centering
%         \includegraphics[width=0.8\textwidth]{figs/example1-a.pdf}
%         \caption{ADL}
%         \label{fig:example1-a}
%     \end{subfigure}
%     \hfill
%     \begin{subfigure}[b]{0.57\textwidth}
%         \centering
%         \includegraphics[width=0.8\textwidth]{figs/example1-b.pdf}
%         \caption{OpenAI/Swarm}
%         \label{fig:example1-b}
%     \end{subfigure}
%     \caption{A comparison of the (partial) implementations of a ``flight change assistant'' in ADL and OpenAI/Swarm.}
%     \label{fig:example1}
% \end{figure*}

\begin{figure*}[t]
  \centering
  \includegraphics[width=\textwidth]{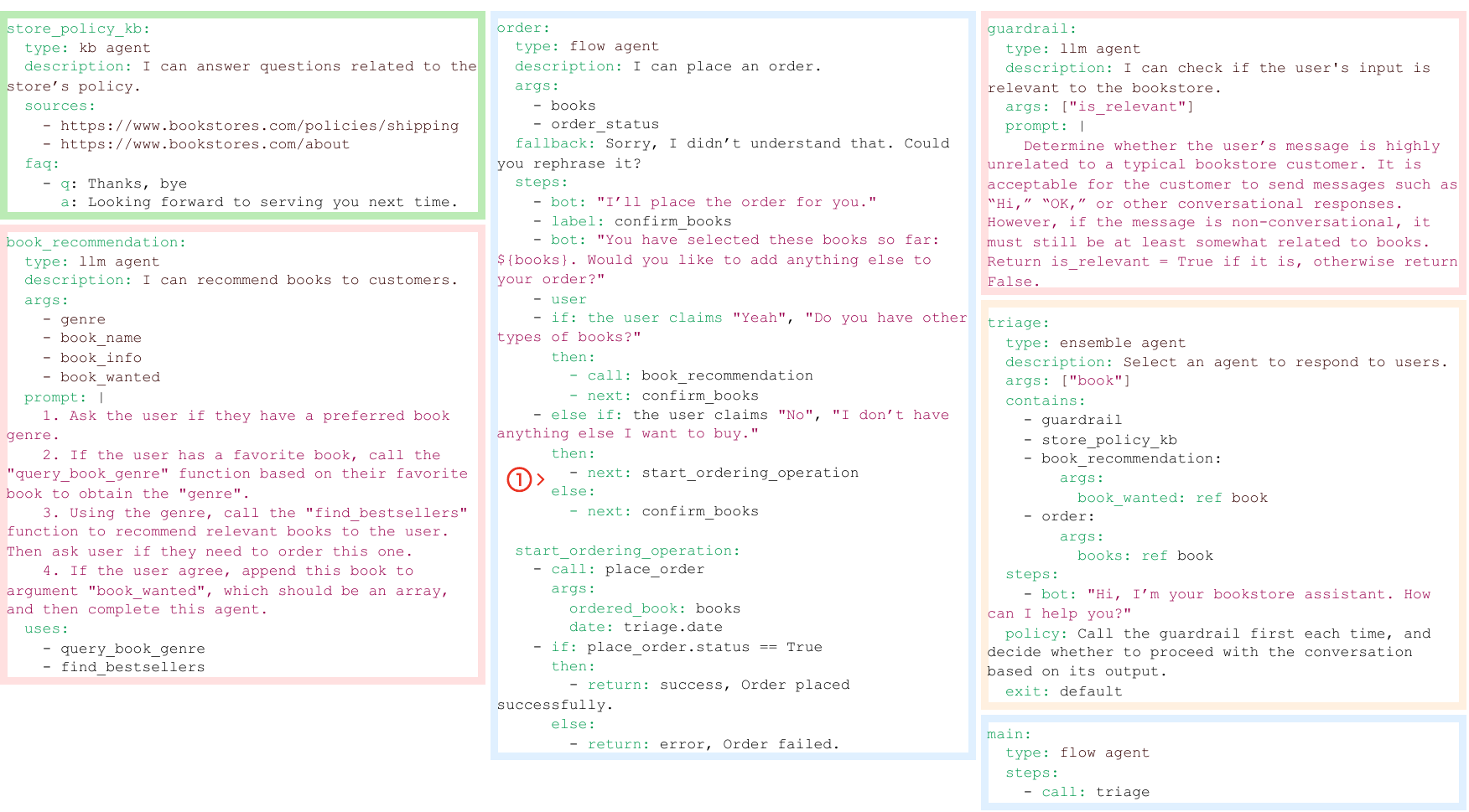}
  \caption{A bookstore chatbot written in ADL}
  \label{bookstore}
\end{figure*}

To address this challenge, we introduce Agent Declarative Language (ADL) that focuses on providing a concise and structured description of agents and their relationships.  The principle behind ADL is to define what agents are and how they interact, rather than how they are implemented and optimized. This level of abstraction leaves system issues to the underlying multi-agent system (MAS), eliminating the need of  manual reprogramming when the MAS changes.

Approaches such as Swarm's minimalist design and CrewAI's use of agent configuration files offer promising directions. ADL takes a bold step further by upgrading these configuration files to agent programs that can be interpreted and executed.  We formulate the design principles of ADL as follows: (1) Make it declarative and modular: Separate logic from execution as much as possible.  (2) Use natural language programming extensively to simplify programming, and (3) Leaving runtime optimization issues to the underlying MAS.  In this way, the performance of ADL programs can grow with the advances of LLMs without revising their code.

We have released MICA, a multi-agent system designed to interpret and execute ADL programs at github \url{https://github.com/Mica-labs/MICA} and received feedback from a few novices that it is much faster to learn and program. In the following sections, we will first introduce the main highlights of ADL and then present two use cases that could demonstrate the unique features of ADL that do not exist in leading agent systems.  The more detailed language specification and examples are given in the Appendices.

\nop{
\begin{table*}[t!]
    \centering
\begin{tabular}{lccccc}
\hline
Framework & Simulation & Tracking & Debugging & Evaluation & Maintainability \\
\hline
Swarm(OpenAI Agents SDK) & - & \checkmark & - & - & - \\
AutoGen & - & \checkmark & \checkmark & - & - \\
CrewAI & - & \checkmark & \checkmark & \checkmark & \checkmark \\
LangGraph & - & \checkmark & - & \checkmark & - \\
\textbf{ADL} & \checkmark & \checkmark & \checkmark & \checkmark & \checkmark \\ 
\hline
\end{tabular}
    \caption{Comparison of ADL and Other Frameworks by Functionality}
    \label{tab:comparison}
\end{table*}
}

\section{Declarative Language}
\nop{Agent Declarative Language (ADL) is a language designed to simplify the creation and management of customer service bots. Essentially, it allows you to describe what you want a bot to achieve (the desired state) without explicitly detailing the steps required (the procedural "how").}  

Agent is a very natural choice for ADL as it supports modular design. Each agent can be tested independently and then combined together for a given task. Figure \ref{bookstore} shows a chatbot written in ADL, which manifests a few basic functions of a bot for bookstore service.  There are four types of agents in ADL: Knowledge base (KB), LLM, Flow, and Ensemble. KB agents handle information retrieval and question-answering tasks, while LLM agents deal with business logic and workflows using natural language. In contrast, flow agents allow traditional control flows through a domain-specific language. An ensemble agent orchestrates other agents. KB agents are atomic meaning they cannot contain or call other agents. All other types of agents can call each other.  The complete language specification of ADL 1.0 is given in Appendix \ref{specification}.

Guardrail agents can be implemented as LLM-based agents and added into an ensemble agent, where they are invoked before and after other agents to detect potentially problematic user inputs or bot outputs. Figure \ref{bookstore} has a relevance guardrail agent to prevent the bot from answering irrelevant user input. 

\nop{The focus of ADL is to let the use declare the business logic and desired outcomes and constraints for each agent.  It hides the underlying implementation complexities as much as possible, making agent creation more accessible to users without extensive programming knowledge.}

ADL supports fast prototyping through LLM agents while leaving step-by-step fine control to flow agents. Meanwhile, ADL enables integration with custom functions, tools, and third-party agents so it can leverage the capability of well established library and routines.  The detailed syntax and grammar are explained in Appendix \ref{language}.

\section{Natural Language Programming}

ADL incorporates elements of natural language, aiming to make it easier for a wider audience to define and customize agents.  In the following, we will showcase what are missing in the existing systems and our solution to these issues. 

\subsection{Interactive Syntax}
Chatbot by nature is interactive: one round of input from user and one round of output from the bot.  However, it is hard to see this from many existing systems. The following   AutoGen~\citep{autogen2023} code shows an agent gets an input from a user and then check if it satisfies a condition or not, it then replies and waits another input from the user.  

\small
\begin{minted}[breaklines=true, breakanywhere=true]{python}
while True:
    user = input("User: ").strip()
    msg = TextMessage(content=user, 
                source="user")
    rsp = await agent.on_messages([msg])
    intent = rsp.chat_message.content
    if intent == "...":
        print("Bot: Continue...")
    else:
        print("Bot: Sorry...")
\end{minted}
\normalsize

From the context, it is not very clear where is the user input from.  In ADL,  we explicitly show this interaction process. 

\small
\begin{minted}[breaklines=true, breakanywhere=true]{yaml}
- user
- if: the user claims "..."
  then:
    - bot: "Continue..."
  else:
    - bot: "Sorry..."
\end{minted}
\normalsize

\subsection{Natural Language Condition}

Most existing systems rely on programming languages to define conversation logic. They handle user intents using either string matching or LLM-based intent checking.  A distinctive feature of ADL is the use of \emph{natural language conditions}, in which user input is directly analyzed by an LLM, improving readability. Consider an example from CrewAI, where the chatbot checks if the user is asking about baggage allowance. 

\small
\begin{minted}[breaklines=true, breakanywhere=true]{python}
query = input("User: ").strip()
prompt = [{"role": "user", "content": "Your task is to determine the user's intent. If the user's intent is related to baggage allowance, return Yes; otherwise, return No."}]
answer = request.post(llm_api, json=prompt)
if "yes" in answer:
    return "There is no free baggage allowance for economy class."
\end{minted}
\normalsize

ADL simplifies it by employing a single natural language conditional statement.\nop{It semantically analyzes user inputs rather than relying solely on literal string matching.}

\small
\begin{minted}[breaklines=true, breakanywhere=true]{yaml}
- user
- if: the user asks anything related to baggage
  then:
    - bot: "There is no free baggage allowance for economy class."

\end{minted}
\normalsize

\subsection{Natural Language Policy}

All agents have a few basic fields to declare, including the task it can handle (description field), its fallback policy (fallback field) and termination condition (exit field) in natural language so that these policies can assist LLMs to select the right agent to handle user requests. 

\small
\begin{minted}{yaml}
base:
  type: <string>
  description: <string> 
  args (optional): <array>
  fallback (optional): <string | agent>
  exit (optional): <string | agent> 
\end{minted}
\normalsize

For example, the exit policy could be ``If there is no input from the user for 30 seconds, please remind the user for an input.  If there is no input for 3 minutes, please say good-bye to the user."  All these policies can be described in natural language and obliterate the need of using lengthy statements to define them.

\nop{
The fallback field describes the fallback policy if the agent cannot
handle the user request. The fallback policy can also assign a specific agent to handle the fallback process. The exit
field illustrates the termination condition of the agent, e.g., if there is no input from the user for a long time.  All these policies can be described in natural language and obliterate the need of using complicated syntax to define these policies. }
 
An ensemble agent can declare a policy when a guardrail agent is called. 

\small
\begin{minted}[breaklines=true, breakanywhere=true]{yaml}
triage:
  type: ensemble agent
  contains: [guardrail_agent, ...]
  policy: call guardrail_agent before selecting an agent everytime.
\end{minted}
\normalsize

\section{Separation of Logic and Optimization}

Using ADL, developers can focus on the desired bot output rather than system performance tuning, which shall be done by the underlying MAS.    

There are many opportunities for performance optimization. For example, there is no need to use the same LLM in processing all kinds of user inputs. It might be critical to leverage state-of-the-art language models for LLM agents and ensemble agents, while KB agents and flow agents can adopt less expensive but fast models. Furthermore, when multiple agents are involved, there are multiple ways to orchestrate them together.  Here we list some of them. 

\begin{itemize}
    \item \textbf{Merging:} Merge two consecutive LLM agents into one-for example, combining a guardrail agent with the following agent, if the guardrail is executed first.
    \item \textbf{First-Success:} Use a round-robin order to iterate through agents until one generates a satisfactory response.
    \item \textbf{Best-of-N:} Execute all agents in parallel and select the best response based on their outputs.
    \item \textbf{Proactive:} Select the most suitable agent based on its description and the user request.
    \item \textbf{Autonomous:} Allow the currently active agent to determine the next agent to invoke.
\end{itemize}

We conducted a comparative analysis of the five orchestration methods described above using the example provided in Figure~\ref{bookstore}.\nop{ The example depicted in Figure~\ref{bookstore} includes a guardrail and four distinct agent types: \textit{kb}, \textit{flow}, \textit{llm}, and \textit{ensemble}.} Initially, each agent type has approximately 100--200 tokens, and this number gradually increases as the conversation progresses.
% \begin{table}[ht]
% \centering
% \begin{tabular}{lc}
% \hline
% \textbf{Agent Name} & \textbf{#Token} \\
% \hline
% store\_policy\_kb (kb agent)         & 192            \\
% book\_recommendation (llm agent)   & 162            \\
% order (flow agent)     & 87           \\
% guardrail (llm agent) & 107     \\
% triage (ensemble agent)       & 104            \\
% \hline
% \end{tabular}
% \caption{Number of Token for each Agent}
% \label{tab:token_num}
% \end{table}

The experiment simulated realistic conversational scenarios: Apart from the initial round, each subsequent dialogue turn represented a shift in user intent, necessitating a transfer from one agent to another. For instance, the statement \textit{"Actually, I'd like to place an order for a book instead."} indicates the user's intent changing from the \textit{book\_recommendation} agent to the \textit{order} agent. 

We evaluated the orchestration methods using two metrics:
\begin{itemize}
\item \textbf{Token Cost}: Defined as the average total number of tokens used in the prompts of all agents invoked during each conversational turn.
\item \textbf{Latency}: Defined as the average elapsed time from user input to bot response.
\end{itemize}

\begin{table}[ht]
\centering
\begin{tabular}{lcc}
\hline
\textbf{Method} & \textbf{Token Cost} & \textbf{Latency} \\
\hline
Merging         & 540       & 2.34s     \\
First-success   & 1732       & 10.39s     \\
Best-of-N       & 1301      & 7.29s     \\
Proactive       & 540      & 3.36s     \\
Autonomous      & 601       & 3.26s     \\
\hline
\end{tabular}
\caption{Comparison of orchestration methods}
\label{tab:method_cost}
\end{table}

We run each dialogue sequence five times and average the results.  All experiments were performed using GPT-4o-mini, and the results are summarized below. Experimental results indicate that the merging method achieves the lowest cost and latency by integrating the guardrail with other agents, thereby reducing the number of required API calls. In contrast, the first-success and best-of-N methods typically involve calling all agents in each conversation round. Particularly, the first-success method incurs the highest latency and cost, as it necessitates an ensemble agent call after each agent invocation to determine if further calls are required. The proactive method differs from merging primarily in that it does not integrate the guardrail with the agents, resulting in similar token costs but slightly higher latency compared to merging. Finally, the autonomous method eliminates the need for ensemble agent calls but incurs a slightly higher token execution cost. Overall, Best-of-N has the highest output quality, but its cost is high.  Due to space constraints, we do not examine more complex scenarios, such as utilizing different LLMs to navigate trade-offs between cost, latency, and output quality. However, ADL offers a flexible abstraction that is not tied to a specific system, making it well suited for future research on system-level optimization.

\section{Case Study: Maintainability}
\label{maintainability}

\nop{show a case if a natural language conditon changes, what modification we need for ADL and the correponding modification we need for Swarm. }

After a customer service bot is deployed, revisions are often required due to initial development oversights or changes in business logic. \nop{For example, customers might become confused by an unclear discount policy during a sale and seek clarification-situations that are unpredictable or may have been overlooked by the company initially. In such cases, rapid fixes or incremental adjustments are required. Existing Python programming frameworks represented by Swarm do not excel in addressing this issue. The following example illustrates this limitation.}The initial implementation of a bookstore chatbot (Figure \ref{bookstore}) handles only scenarios in which customers wish to add additional books to their orders. However, in practice, customers may inquire about available discounts. Suppose that the management team would like to add a new discount interaction flow. Implementing this enhancement involves adding a few lines of code, whereas achieving the same functionality in Swarm requires more extensive modification if step-by-step control is needed. 
% \begin{figure*}[t!]
%   \centering
%   \includegraphics[width=\textwidth]{figs/update-adl.pdf}
%   \caption{ADL enables easy adaptation to dialogue policy changes with minimal modifications.}
%   \label{update-adl}
% \end{figure*}
In the location of \textcircled{1}, Figure \ref{bookstore}, we can add an additional branch statement. \nop{This capability enables rapid chatbot logic updates through only a few minor adjustments.}

\small
\begin{minted}[breaklines=true, breakanywhere=true]{yaml}
- else if: the user asks for any discount
  then:
    - bot: "Here's a special discount for you..."
\end{minted}
\normalsize

% \begin{figure*}[t!]
%   \centering
%   \includegraphics[width=\textwidth]{figs/update-swarm.pdf}
%   \caption{Modifications needed for updating identical chatbot logic in Swarm}
%   \label{update-adl}
% \end{figure*}

\nop{In contrast, Swarm does not offer similar precise control, necessitating reliance on Python code for such logic management. Consequently, logic changes in Swarm require more complex considerations and code modifications compared to ADL. To determine subsequent workflows based on user intent, an \emph{intent classifier agent} must be explicitly defined and maintained. Each time the set of intents changes, the definition of this agent must be updated correspondingly. Furthermore, changes in conversation logic often introduce additional internal state variables. For example, in this scenario, an additional variable such as \texttt{subflow\_tries} is required to count instances of user hesitation. As scenarios grow increasingly complex, the resulting code tends to suffer further deterioration in readability and maintainability.}

In contrast, logic changes in Python require more complex syntax and code modifications. To determine subsequent workflows based on user intent, an \emph{intent classifier agent} must be explicitly defined and maintained. Each time the set of intents changes, the definition of this agent must be updated correspondingly. Furthermore, changes in conversation logic often introduce additional internal state variables. As scenarios grow increasingly complex, the resulting code tends to suffer further deterioration in readability and maintainability.

\small
\begin{minted}[breaklines=true, breakanywhere=true]{python}
    user_msg = input().strip()
    intent = call_agent(
        new_intent_classifier, # update intent list
        user_msg)
    if intent == "ask_discount":
        while True: # discount subflow
            ...     # offer a discount
            if ...: # termination condition
                break
    ...
\end{minted}
\normalsize

\section{Case Study: Debugging Assistance}
In practice, creating a service chatbot rarely proceeds without issues. Much like traditional software development, the process is prone to the emergence of various bugs, making effective debugging an essential aspect of chatbot engineering. Most existing MAS frameworks log the conversation history, allowing developers to manually inspect dialogue flows, identify anomalies, and trace issues back to specific prompts or logic components.However, this manual approach can be time-consuming and inefficient. As prompts grow in complexity and agent interactions become increasingly intricate, locating the source of bugs quickly becomes a significant bottleneck. Unlike humans, LLMs 
are particularly adept at parsing and analyzing extensive textual data, offering a compelling opportunity to leverage their capabilities for more efficient debugging.

Consider the example presented in Section \ref{maintainability}. The natural language flow in ADL appears more comprehensible than the hybrid form that intermixes natural language with Python programming. We argue that ADL offers a distinct advantage of leveraging LLMs for debugging assistance. While a comprehensive evaluation is beyond the scope of this work, the following section provides an illustrative example.  \nop{ demonstrating how LLMs can effectively support the debugging of ADL programs. }

\subsection{Retail Banking Bot} We implemented a \textit{Retail Banking Bot} using ADL and Swarm. The bot consists of eight distinct agents, each responsible for a specific functionality: FAQ handling, money transfer, adding payees, deleting payees, listing all payees, balance inquiries, and blocking cards. The complete program is provided in Appendix \ref{banking}. We intentionally insert an infinite loop bug in the implementation. 

\nop{
To simulate realistic development errors, we discuss three distinct types of bugs:

\begin{enumerate}
\item \textbf{Explicit Prompt Errors in Individual Agents:} These errors occur within the definition of a single agent. Examples include incorrect conditional logic, erroneous tool calls, or incomplete logic implementations, resulting in unexpected behavior during testing.

\item \textbf{Implicit Errors Caused by Inter-Agent Dependencies:} Such errors occur when one agent mistakenly modifies variables or states, negatively impacting subsequent agents' behaviors.

\item \textbf{Infinite Loops:} These occur when two or more agents repeatedly invoke each other, leading to frequent context switching and preventing conversation progression.

\end{enumerate}

In our experiments, erroneous implementations of Swarm and ADL code were provided to an LLM, along with corresponding conversation histories and expected correct behaviors, to evaluate the LLM's capability to automatically debug by pinpointing errors. Details of these experiments are described in the following sections.
}

\subsection{Debugging Infinite Loops}

Infinite loop is a common issue in multi-agent systems, particularly when multiple agents interact and invoke each other. Due to the use of natural language programming in chatbot development, such errors are difficult to detect with traditional debugging tools. In existing frameworks, these errors can only be identified through replay and agent by agent checking. The cost of finding such bugs is high, as they require manual inspection of dialogues covering all potential interaction paths. Consequently, employing LLMs for static analysis could improve debugging efficiency.

We simulated a scenario illustrating a possible infinite loop involving two agents: one responsible for transferring money and the other for adding a payee. During the money transfer process, the user must specify a payee, prompting the invocation of the \texttt{add\_payee} agent to ensure the payee's validity. 

\small
\begin{minted}[breaklines=true, breakanywhere=true]{yaml}

transfer_money:
  prompt: |
    ...
    4. ...enter the add_payee agent...
    (Appendex D.1 transfer_money agent)
    ...
\end{minted}
\normalsize

However, if the \texttt{add\_payee} agent incorrectly prompts the user regarding money transfer and the user responds affirmatively, an infinite loop occurs, continuously cycling between the two agents and preventing the transfer process from completing.

\small
\begin{minted}[breaklines=true, breakanywhere=true]{yaml}
add_payee:
  prompt: |
    ...
    4. Ask the user if they need to make a transfer; if so, enter the transfer_money agent.
    (Appendex D.1 add_payee agent)
\end{minted}
\normalsize

We tested LLMs' ability to detect infinite loops under zero-shot conditions: Submit the ADL or Swarm code and ask GPT-4o if there is an infinite loop in the code (Appendix \ref{prompt}). The detection results are summarized in Table \ref{tab:debug}. In the Swarm implementation, the LLM incorrectly identified a loop between the \texttt{triage} and \texttt{greeting} agents. In practice, this loop is impossible since the \texttt{triage} agent can determine whether the \texttt{greeting} agent has been previously invoked, thus preventing loops. Conversely, the LLM accurately detected the infinite loop in the ADL implementation.
\begin{table*}[htbp]
\centering
\begin{tabular}{lcc}
\hline
\textbf{Framework} & \textbf{Detected Loop}   & \textbf{Actual Presence of Loop}\\
\hline
Swarm           &  triage $\to$ greeting $\to$ triage $\to$ ...                     & \ding{55} \\
ADL             &  transfer\_money $\to$ add\_payee $\to$ transfer\_money $\to$ ... & \checkmark \\

\hline
\end{tabular}
\caption{Infinite loop detection for the same service bot implemented in Swarm and ADL (Appendix \ref{banking})}
\label{tab:debug}
\end{table*}

\begin{table*}[htbp]
\centering
\begin{tabularx}{\textwidth}{X}

\hline
There is a potential infinite loop involving these agents:
1.	\textbf{transfer\_to\_triage()} is included in the functions list of many agents (e.g., greeting, add\_payee, block\_card, etc.).

2.	The \textbf{triage\_agent} itself can transfer to any of those agents again, including \textbf{greeting} via \textbf{transfer\_to\_greeting}.
This creates an infinite loop of redirection between agents if there's no stop condition or terminal action in place. Similar loops can happen with any agent that calls transfer\_to\_triage and is itself callable from triage\_agent. \\
\hline

\end{tabularx}
\caption{LLM reasoning over the Swarm code (partial)}
\label{tab:reasoning}
\end{table*}

Table \ref{tab:reasoning} shows the LLM's chain-of-thought reasoning for the Swarm code. There are misinterpretations related to function execution-specifically, the incorrect inference that agent termination triggered additional function calls, resulting in unintended loops. In contrast, ADL definitions allow the LLM to focus more on agent functionality, reducing the likelihood of such misinterpretations. Because Swarm integrates natural language prompts with Python code, it poses challenges for LLMs in understanding the intended semantics. We hypothesize that this difficulty may be exacerbated when fine-grained control flow logic is implemented using Python + prompts in Swarm vs. flow agents in ADL. Further investigation will be pursued in future work.

\section{Related Work}
Multi-agent systems provide an effective solution for handling complex tasks~\citep{dorri2018multi}. Negotiation and interaction among multi-agents has been studied extensively, e.g., \citet{beer1999negotiation}. \citet{ferber1998meta} proposed constructing a meta-model to represent all kinds of tasks, and other researchers have explored how to build a unified framework for developing multi-agent systems tailored to various domains \citep{bellifemine2001developing}. \citet{zolitschka2020novel} considered cooperation among agents and proposed the use of middle agents as an orchestration layer, responsible for integrating responses from all other chatbot agents.\nop{A substantial body of work demonstrated the effectiveness of multi-agent systems and explored the relationships between different agents. \citep{mcarthur2007multi,mcarthur2007multi2,catterson2012practical} However, these approaches are often limited by the complexity of agent design. Building a high-performance multi-agent system typically requires extensive prior knowledge and domain expertise, as well as significant programming experience. This limitation hinders the widespread application of multi-agent systems across different domains.} 

Various agent programming languages are available, such as AgentSpeak~\citep{agentspeak} and JADE~\citep{bellifemine2001developing}.\nop{In the past, there was an international workshop devoted for Declarative Agent Languages and Technologies.} Among these efforts,  GOAL~\citep{GOAL} distinguished itself with the concept of a declarative goal. \nop{which describes what an agent wants to achieve, not how to achieve it.}

With the advancement of LLMs, agents have become more capable of handling diverse tasks.\nop{ while significantly reducing the complexity of their construction.} LLM-based multi-agent systems has attracted widespread attention \citep{kannan2024smart, obata2024lip, talebirad2023multi, sun2024llm}. 
These studies aimed to find a unified and convenient approach to agent design. \citet{autogen2023} introduced a framework for customizing agents and enabling their cooperation, offering developers a relatively simple implementation pathway. However, it does not distinguish between declaration and implementation.  Similar efforts can also be found in numerous agent frameworks such as LangGraph~\citep{LangGraph}, CrewAI~\citep{crewAI}, and Swarm~\citep{Swarm}, which provide high flexibility for constructing agents in general applications.  There are great benefits to mingle  agent declaration with Python codes as it can leverage Python library to complete many functions of agents.  However, for customer service chatbots, a simpler approach like ADL might be feasible, where definitions of agents are decoupled from Python programming. 

\citet{trirat2024automl} attempted to extract different agents directly from user-provided natural language descriptions of tasks. \citet{rebedea-etal-2023-nemo} investigated the feasibility of defining chatbots entirely using natural language, making the system more accessible to developers. However, their design requires more elaborate definitions when handling increasingly complex logic. \nop{In contrast, ADL provides a structured and simplified framework for defining chatbot behavior, ensuring a more intuitive, modular, and maintainable design process.}

\section{Conclusion}
ADL is a simple, declarative language designed to facilitate a clear definition of multi-agent systems for task-oriented chatbots. By leveraging declarative programming, ADL separates agent task specification from system implementation and optimization, making the program easier to understand and maintain.   ADL is fully operational and available as an open-source package at \url{https://github.com/Mica-labs/MICA}. Further research shall be done to standardize interaction patterns among multiple agents and explore declarative languages for more complicated scenarios targeted by the leading multiple agent systems, as well as system optimization of underlying MASs that could run ADL programs. 

\section*{Limitations}
Our study has several limitations. First, due to space constraints, we did not evaluate the performance of different orchestration methods in more complex scenarios. Our comparisons are limited to token cost and latency; further experiments using ADL for system-level optimization remain necessary.

Second, our research on automatic debugging focused exclusively on infinite loops, a common yet singular development issue. Further work should classify and investigate additional development problems and the effectiveness of LLMs for automatic debugging. Additionally, as agent applications become more complex, requiring finer-grained control over conversation logic, the performance gap between declarative definitions and programming definitions as understood by LLMs may widen, necessitating further experimental validation.

\bibliography{works}

\newpage
\appendix
\onecolumn
\section{ADL 1.0 Language Specification}
\label{specification}

ADL 1.0 is a descriptive language that conforms to the YAML specification. A complete ADL program consists of multiple \texttt{agent} objects. The syntax rules of ADL written in extended Backus-Naur form (EBNF) are as follows:

\small
\begin{align*}
\textbf{adl} \quad &::= \texttt{main:} \; \textbf{agent}  \; \{\, \textbf{agent\_name}\texttt{:}\; \textbf{agent} \,\}  \; [\;\texttt{tools:} \; \textbf{tool\_name}+ \,] \\[1ex]
\textbf{agent\_name} \quad &::= \textbf{string} \\[1ex]
\textbf{agent} \quad &::= \textbf{kb\_agent} \; |\; \textbf{llm\_agent} \; |\;\textbf{flow\_agent} \; |\;\textbf{ensemble\_agent} \\[1ex]
\textbf{tool\_name} \quad &::= \textbf{string} \\[1ex]
\end{align*}
\normalsize

A valid ADL program must contain a \texttt{main} agent, which serves as the entry point of the program. It follows the same structure as other agents but has a fixed name, \texttt{main}. It may include multiple \texttt{agent}s to define agents with different functionalities. Additionally, the program may include a  \texttt{tools} block, which refers to python functions that are going to be called in agents.

Each \texttt{agent} contains a set of common attributes, collectively referred to as the \texttt{agent\_header}. The \texttt{agent\_header} includes a mandatory \texttt{description} field, which specifies the primary functionality of the agent. This field serves as documentation to help other agents understand the role of the given agent. The \texttt{args} field may also be included to list all arguments associated with the agent.

Both \texttt{fallback} and \texttt{exit} are optional fields used to handle edge cases during agent execution. The \texttt{fallback} field specifies the strategy to be invoked when the agent does not know what to do with the current  user request.  It can either be the name of another agent responsible for fallback handling or a natural language description outlining the fallback strategy. The \texttt{exit} field defines the strategy to be executed once the agent completes its task. Its structure is identical to that of the \texttt{fallback} field. 

\small
\begin{align*}
\textbf{agent\_header} \quad &::= \textbf{description} \;  [\; \textbf{args} \;] \;  [\; \textbf{fallback} \;] \;  [\; \textbf{exit} \;] \\[1ex]
\textbf{description} \quad &::= \texttt{description:} \;  \textbf{string} \\[1ex]
\textbf{args} \quad &::= \texttt{args:} \; \textbf{arg\_name}+ \\[1ex]
\textbf{fallback} \quad &::= \texttt{fallback:} \; \textbf{agent\_name} \; | \; \textbf{string}\\[1ex]
\textbf{exit} \quad &::= \texttt{exit:} \; \textbf{agent\_name} \; | \;\textbf{string}\\[1ex]
\end{align*}
\normalsize

 Within each \texttt{agent}, its body differs according to the agent type.  The \texttt{kb agent} includes parameters relevant to knowledge-based agents. Among these, \texttt{sources} is an optional field that lists various files or links-such as plain text documents, PDFs, CSVs, or URLs-that contain domain knowledge. The \texttt{faq} field is also optional; each item in the \texttt{faq\_block} must follow the structure \texttt{\{'q': string, 'a': string\}}.

In the \texttt{llm agent}, the \texttt{prompt} field is mandatory, while the \texttt{uses} field is an optional list of function names that the \texttt{llm agent} may utilize. It also has an optional \texttt{step\_block} that could contain  initialization opertions. 

The \texttt{flow agent} must include a \texttt{step\_block}, consisting of multiple step statements. Its detail will be given in the next section. Subflows, if present, share the same structure as the \texttt{steps} field.

The \texttt{ensemble agent} includes a \texttt{contains} field, which is a list of agents managed by the \texttt{ensemble\_agent}. The \texttt{args} keyword represents all the arguments specific to the ensemble agent. The \texttt{ref} field needs to be set when the agent is allowed to modify the values of arguments within the ensemble agent. Both the \texttt{prompt} and \texttt{steps} fields are optional. The \texttt{prompt} field may specify the agent selection strategy employed by the \texttt{ensemble agent}, while the \texttt{steps} field defines the initialization steps of the ensemble agent, sharing the same \texttt{step\_block} structure as used in the \texttt{flow agent}.

{\small
\begin{align*}
\textbf{kb\_agent} \quad &::= \textbf{agent\_header}, \; \texttt{type:kb agent}, \; [ \; \texttt{sources:} \;  \textbf{string}+ \; ], \; [ \; \texttt{faq:} \;  \textbf{faq\_block}+ \; ]\\[1ex]
\textbf{llm\_agent} \quad &::= \textbf{agent\_header}, \; \texttt{type:llm agent},\; \texttt{prompt:} \; \textbf{string}, \; [ \; \texttt{uses:} \; \textbf{function\_name}+ \; ], \; [ \; \textbf{step\_block} \;] \\[1ex]
\textbf{flow\_agent} \quad &::= \textbf{agent\_header}, \;  \texttt{type:flow agent},\;  \textbf{step\_block}  \\[1ex]
\textbf{ensemble\_agent} \quad &::= \textbf{agent\_header}, \;  \texttt{type:ensemble agent},\; \texttt{contains:} \;  \textbf{invoke\_agent}+ \; , \; [ \; \texttt{prompt:} \; \textbf{string}\; ], \; \\[1ex] 
\quad &\quad \quad [ \; \textbf{step\_block} \; ] \\[1ex]
\textbf{faq\_block} \quad &::= \texttt{q:} \; \textbf{string}, \; \texttt{a:} \; \textbf{string}  \\[1ex]
\textbf{invoke\_agent}  \quad &::= \textbf{agent\_name}\texttt{:}\; [\; \texttt{args:}\;(\; \textbf{arg\_name}: \; [\; \texttt{ref} \;] \; \textbf{arg\_name} \;)+  \;] \\[1ex]
\end{align*}
\normalsize}

The \texttt{step\_block} is ADL's mechanism for flow control. It could contain different types of step statements.

{\small
\begin{align*}
\textbf{step\_block} \quad &::= \texttt{steps:} \; \textbf{step}+, \; \{ \;\textbf{subflow\_name} \; \texttt{:} \; \textbf{step}+ \; \}\\[1ex]
\textbf{subflow\_name} \quad &::= \textbf{string} \\[1ex]
\textbf{step} \quad &::= \texttt{user}\\[1ex]
\quad &\quad \; | \; \texttt{bot:}  \; \textbf{string} \\[1ex]
\quad &\quad \; | \; \texttt{set:} \; ( \; \textbf{arg\_path}:  \; \textbf{string} \;|\; \textbf{arg\_path} \; )+ \\[1ex]
\quad &\quad \; | \; \texttt{label:} \; \; \textbf{label} \; \\[1ex]
\quad &\quad \; | \; \texttt{next:} \; \; \textbf{label} \; | \; \textbf{subflow\_name}, \; [\; \texttt{tries:} \; \textbf{number}\;] \\[1ex]
\quad &\quad \; | \; \texttt{call:} \; \; \textbf{agent\_name} \; | \; \textbf{function\_name}, \; [\;\texttt{args:} \; (\; \textbf{arg\_name}:  \; \textbf{string} \;|\; \textbf{arg\_path} \; )+ \;] \\[1ex]
\quad &\quad \; | \; \textbf{condition\_step}\\[1ex]
\quad &\quad \; | \; \texttt{return:} \; \textbf{string}\\[1ex]
\textbf{condition\_step} \quad &::= \texttt{if:} \; \textbf{conditional\_expression} \; \texttt{then:} \; \textbf{step}+ ,\; [ \; \textbf{else\_clause} \;]\\[1ex]
\textbf{else\_clause} \quad &::= \texttt{else:} \; \textbf{step}+  \; |\; \texttt{else if:} \; \textbf{conditional\_expression} \; \texttt{then:} \; \textbf{step}+ , \;  [ \; \textbf{else\_clause} \; ] \\[1ex]
\textbf{arg\_path} \quad &::= [\;\textbf{agent\_name}.\;] \; \textbf{arg\_name} \\[1ex]
\textbf{arg\_name} \quad &::= \textbf{string} \\[1ex]
\textbf{function\_name} \quad &::= \textbf{string} \\[1ex]
\textbf{conditional\_expression} \quad &::= \textbf{string} \\[1ex]
\end{align*}
\normalsize}

The \texttt{user} step serves as a placeholder indicating that the agent should wait for a user response, while the \texttt{bot} step represents the agent's reply. The \texttt{label} can appear anywhere within the \texttt{steps}, and the \texttt{next} keyword can be used to jump to any label or subflow within the same \texttt{flow agent}.

The \texttt{call} field indicates that the agent should invoke another agent or function. The \texttt{args} attribute specifies the parameter mapping required for the call, where the key refers to the parameter name in the target agent or function, and the value indicates either a static value or a path to the source of the value.

The control flow constructs \texttt{if}, \texttt{else if}, and \texttt{else} are used for conditional branching. Both \texttt{if} and \texttt{else if} are followed by a condition expression, which can represent either a user intent or a parameter-based condition. If the expression evaluates to be true, the steps under the \texttt{then} block are executed; otherwise, if an \texttt{else} block exists, its steps are executed. We do not further elaborate on the syntax and grammar of condition expressions, as it is anticipated that conditions will more frequently be expressed in natural language.

The \texttt{return} statement indicates that the agent should terminate immediately. An agent may terminate in one of two status: \texttt{success} or \texttt{error}. Each of these can be followed by a specific \textbf{message} to indicate the outcome.

Some basic elements are shown below.

\nop{
\textbf{string\_list} \quad &::= (\; \texttt{-} \; \textbf{string} \; )\; \{\; \texttt{-} \; \textbf{string} \,\} \\[1ex]
}
\small
\begin{align*}
\textbf{string} \quad &::= \textbf{letter} \; \{\, \textbf{letter} \; | \; \textbf{digit} \; | \; \texttt{\_}  \; | \;  \texttt{.}\,\} \\[1ex]
\textbf{number} \quad &::= \;\textbf{digit}+ \; \\[1ex]
\textbf{letter} \quad &::= \texttt{A} \; | \; \texttt{B} \; | \; \cdots \; | \; \texttt{Z} \; | \; \texttt{a} \; | \; \cdots \; | \; \texttt{z} \\[1ex]
\textbf{digit} \quad &::= \texttt{0} \; | \; \texttt{1} \; | \; \cdots \; | \; \texttt{9} \\[1ex]
\end{align*}
\normalsize

ADL will continue to evolve. Its interface with tool calls, third-party agents, and external resources-via protocols such as the Model Context Protocol (MCP) \citep{Anthropic2024MCP}-needs to be further specified. Nonetheless, the preliminary specification outlined above reflects the core design principle of maintaining simplicity in agent programming.

\section{Language}
\label{language}
The core of ADL is about agents. There are four types of agents in ADL: Knowledge base (KB), LLM, Flow, and Ensemble Agent. KB agents handle information retrieval and question-answering tasks, while LLM agents deal with business logic and workflows using natural language. In contrast, flow agents allow traditional control flows through a domain-specific language. An ensemble agent orchestrates other agents. KB agents are atomic meaning they cannot contain or call other agents. All other types of agents can call each other.  The complete language specification of ADL 1.0 is given in Appendix \ref{specification}.
% \begin{figure}[t]
% \begin{center}
% %\framebox[4.0in]{$\;$}
% \includegraphics[width=0.3\textwidth]{type_of_agent}
% \end{center}
% \caption{The type of agents.}
% \end{figure}
% Figure. the types of agents
\subsection{Base Agent}
All agents have a few basic fields. 
\small
\begin{minted}{yaml}
base:
  type: <string>
  description: <string> 
  args (optional): <array>
  fallback (optional): <string | agent>
  exit (optional): <string | agent> 
\end{minted}
\normalsize

The \textbf{type} field defines the type of agents. It can choose from four values: "kb agent," "llm agent," "flow agent," and "ensemble agent." The \textbf{description} field provides a general description of what the agent is about. It can assist LLMs to select the right agent to handle user requests. The \textbf{fallback} field describes the fallback policy if the agent cannot handle the user request. The fallback policy can also assign a specific agent to handle the fallback process.  The \textbf{exit} field illustrates the termination condition of the agent, e.g., if there is no input from the user for a long time.  

Each agent will have a body to explain its logic, which will be described in a different way depending on the type of agent.

\begin{figure*}[t!]
    \centering
    \includegraphics[width=0.8\textwidth]{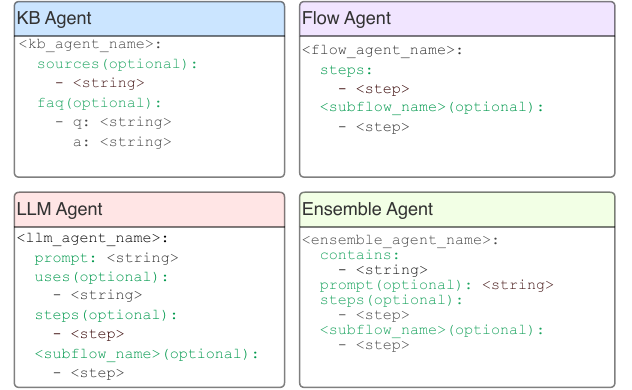}
    \caption{Body of different types of agents}
    \label{fig:agents}
\end{figure*}

\subsection{KB Agent}
KB Agent is designed to perform information retrieval and question answering. Given an information source(s), a KB agent automatically searches for relevant information and returns the answer.  There could be various kinds of sources including \texttt{.pdf}, \texttt{.doc}, \texttt{.txt}, \texttt{.csv}, url, etc.  The \textbf{faq} field stores frequently asked questions along with their corresponding answers, formatted according to the YAML specification.

% \begin{minted}{yaml}
% kb_agent_name:
%     type: kb agent
%     source: <string>
%     faq: <array>
%         question: answer
% \end{minted}

\subsection{LLM Agent}
LLM Agent encodes domain knowledge and constraints through prompt programming. Additionally, an LLM agent can leverage tools declared in the \textbf{uses} clause to call functions written in Python. You are required to provide a brief introduction (the \textbf{description} field), detailed instructions (the \textbf{prompt} field), and any relevant information, such as the names of tools used within the LLM agent. The \textbf{uses} field lists all the function names referenced in the agent's prompt. These functions must be defined in a separate Python file.\nop{When arguments need to be collected or specific functions need to be executed, the agent will automatically handle these components by checking functions in the Python file.} If a function needs to be invoked during the process, the \textbf{prompt} field should indicate when and which function to call. The \textbf{args} field is included when certain states or arguments must be collected or recorded. If the agent determines that it cannot respond adequately based on what is declared in its prompt, it will deactivate itself and transfer control back to other agent.  An LLM agent might include a few initialization steps in the step block which will be executed before the prompt takes over. 

% \begin{minted}{yaml}
% llm_agent_name:
%     type: llm agent
%     description: <string>
%     args: <array>
%     prompt: <string>
%     uses: <array>
% \end{minted}

\nop{

\textbf{Schema}. A typical LLM agent in Figure~\ref{fig:agents} consists of the following attributes. The \textbf{description} field provides a concise explanation of the agent's functionality, serving as a reference for other agents that need to understand its purpose. The \textbf{prompt} field specifies how the agent executes its tasks. If a function needs to be invoked during the process, the prompt should indicate when and which function to call. The \textbf{args} field is included when certain states or arguments must be collected or recorded. Lastly, the \textbf{uses} field lists all function names referenced in the agent's prompt. These functions must be defined in a separate Python file.
}

\subsection{Flow Agent}
Flow Agent offers flow control similar to traditional programming languages but using a YAML form. Flow agents are less declarative.   It ensures that all interactions between the bot and the user  strictly follow the predefined flow. Additionally, it includes a fallback mechanism: when unexpected situations arise, it automatically triggers the fallback mechanism, deactivates itself, and transfers control back to other agents.  The syntax of flow agent shares similarity with Google's Workflow~\citep{googleworkflow}. The flow agent starts with a list of \textbf{steps}, where all logic is explicitly defined.

% \begin{minted}{yaml}
% flow_agent_name:
%     type: flow agent
%     description: <string>
%     args: <array>
%     steps: <array>
%     fallback: <string> | <object>
% \end{minted}

\nop{
\textbf{Schema}. The \textbf{description} keyword is similar to that of an LLM agent; it provides a concise yet informative description of the agent, enabling other agents to recognize its function. The \textbf{args} field, also akin to that in an LLM agent, is a list of all states that need to be recorded. The \textbf{steps} field constitutes the body of the Flow Agent, where all logic is explicitly defined. The \textbf{fallback} field specifies the fallback policy or the name of a specific LLM agent or Flow Agent designed to handle fallback situations; if not provided, the flow will terminate immediately.
}

\subsubsection{Steps}
The flow steps consists of \textbf{user} step, \textbf{bot} step and a set of other statements.  Intuitively, a \textbf{user} step acts as a placeholder that waits for user input. It serves as a marker indicating that all preceding steps have been executed and the system should now wait for the user to respond.  In contrast, a \textbf{bot} step defines what the bot will say. Each bot step represents a single chatbot response. 

\small
\begin{minted}[escapeinside=@@]{yaml}
<flow_agent_name>:
  steps:
    - bot: "hello, world"
    - @\textcolor{keycolor}{\textbf{user}}@
    - call: <agent_name | function_name>
      args: 
        - <arg_name>: <value | arg_path>
\end{minted}
\normalsize

\subsubsection{Call}
\nop{A key feature of the Flow Agent is its ability to execute different agents sequentially according to a predefined order. To enable this functionality, we introduce the \textbf{call} keyword.

1) \textbf{Tool}: Used to interact with external services.
2) \textbf{LLM Agent}: Executes a specific task and then returns to the current Flow Agent.
3) \textbf{Flow Agent}: Switches execution to another Flow Agent, returning to the current agent upon completion.
}

The \textbf{call} keyword can invoke functions and agents. The required parameters can be specified by the \textbf{args} keyword.  The arguments in the callee agent/function can be referred by \texttt{[agent\_name].<arg\_name>} in the caller.  
\nop{
In this example, the \textbf{args} field defines a mapping where the keys correspond to argument names expected by the called object, and the values are either known argument names or explicitly assigned values. As previously introduced, arguments can be retrieved using the format \texttt{[agent\_name].<arg\_name>}, allowing access to arguments from any agent or tool.

Additionally, the bot can output the value of any argument from any agent. By using the format \texttt{\$\{[agent\_name].<arg\_name>\}}, the argument \texttt{arg\_name} from the specified \texttt{agent\_name} will be automatically replaced with its value during output. 
If \texttt{agent\_name} is omitted, the reference defaults to the argument within the current agent.

Within the \textbf{steps} attribute of the Flow Agent, our language supports typical programming constructs such as sequential execution, conditional branching, loops, and jumps. The design strictly follows the YAML specification.
}

\subsubsection{Set}
\nop{Different arguments can also exist within a Flow Agent. 

For instance, consider two agents responsible for order placement and delivery. The order placement agent has an argument called \texttt{billing address}, while the delivery agent has an argument called \texttt{delivery address}. To reduce redundant input, the value of \texttt{billing address} can be manually passed to the delivery agent's \texttt{delivery address} field.
Arguments can also be manually set to any value, including empty values, boolean values, strings, arrays, or any object that conforms to the YAML schema.

Generally, an agent automatically extracts the corresponding arguments from user input.} There are cases where arguments need to be manually assigned. Here is an abridged example (we omitted the code that gets a user's permission to use her billing address as delivery address). 
\small
\begin{minted}{yaml}
delivery:
  - call: shopping_agent
  - bot: "I got your billing address: ${shopping_agent.billing_address}."
  - set:
      delivery_address: shopping_agent.billing_address
\end{minted}
\normalsize

\subsubsection{Subflow}
A subflow is a code block in a flow that can be called inside the flow via the \textbf{next} keyword.
\nop{
It follows this structure: a Flow Agent must contain a \texttt{main} subflow, while other subflows can be referenced as callable labels. Elements within a subflow must be defined as a list.
}

\small
\begin{minted}{yaml}
<flow_agent_name>:
  steps:
    - bot: "This is the main flow."
    - next: a_subflow
  a_subflow:
    - bot: "This is a subflow."
\end{minted}
\normalsize

\subsubsection{Condition}
Conditional evaluation plays a critical role in branching and looping. In a conversation, there are two primary scenarios where conditional evaluation alters the dialogue flow: 
(1) Entering different branches based on the user's intent.
(2) Entering different branches based on an expression of certain arguments.
To achieve this, we designed the \textbf{condition} statement. 

\small
\begin{minted}{yaml}
steps:
  - if: the user claims "<user_intent>"
    then:
      - <step>
      - return: <"success" | "error">, <msg: string>
  - else if: <conditional_expression: string>
    then:
      - <step>
    else:
      - <step>
\end{minted}
\normalsize
The \textbf{if} and \textbf{else if} clauses have user intent checked or conditional expression evaluated. The supported expression formats are shown in Table~\ref{tab:expression}. \nop{When a condition evaluates to \texttt{True}, the corresponding statements inside the \textbf{then} block are executed. If the condition evaluates to \texttt{False}, the statements within the \textbf{else} block are executed.} Conditions can also be described by natural language and let an LLM evaluate if the condition is true or false.

\nop{
This unified design facilitates future integration with LLM-based control. Since LLMs excel at understanding conversational intent, our long-term goal is to enable LLMs to take full responsibility for condition evaluation.
}

\begin{table}[h]
    \centering
    \begin{tabular}{ll}
    \toprule
    \textbf{Condition Type} & \textbf{Syntax} \\
    \midrule
    Checking for \texttt{None} values       & \texttt{<arg\_path> == None} or \texttt{<arg\_path> != None} \\
    Checking for a specific value           & \texttt{<arg\_path> == <value>} \\
    Checking if a value matches a pattern   & \texttt{re.match(<regular expression>, <arg\_path>)} \\
    Combining multiple conditions           &  \texttt{and}, \texttt{or} \\
    \bottomrule
    \end{tabular}
    \caption{Condition Checks in ADL}
    \label{tab:expression}
\end{table}

Unless a \textbf{return} statement is encountered, the flow will continue until it completes. The \textbf{return} keyword indicates the termination of the flow. It returns one of the two statuses: \texttt{success} and \texttt{error}, followed by a custom message to the caller. 

\subsubsection{Loop}
In certain scenarios, it is necessary to return to a specific location of a flow to repeat a few steps.  To achieve this, the \textbf{label} keyword is used for positioning, while \textbf{next} specifies the target location. ADL implements looping behavior through the \textbf{tries} attribute. The \textbf{tries} attribute is optional and defines the maximum number of times to repeat. If omitted, it defaults to unlimited retries.  We leave more structured loop statements for future extension. 

\small
\begin{minted}{yaml}
steps:
  - label: retry
  - bot: "..."
  - next: retry
    tries: 3
\end{minted}
\normalsize
This example simulates a scenario in which a customer is given three attempts to correctly enter a verification code.

The \textbf{label} keyword designates a specific location within a flow. Each label must be unique within the same flow to prevent conflicts and cannot share a name with any subflow. The \textbf{next} keyword specifies the target location, which can be either a label or a subflow.

\subsection{Ensemble Agent}
Ensemble agent determines which agent should respond given a user input and a conversation context. In a multi-agent system, a scheduling strategy is required to manage who responds  to what.  The ensemble agent fulfills this role. Many other frameworks incorporate a similar agent, such as ``orchestrator" in AWS's MAO framework and ``Triage" in OpenAI/Swarm.  We chose the term ``Ensemble Agent" because, in our framework, agents can be peer-led without a central orchestrator.  ADL  allows different coordinating  strategies, which actually can be described in the \textbf{prompt} field using natural language.  \nop{rather than following a strict hierarchy. There is no absolute leader; instead, agents operate independently and equally, with the ability to invoke and collaborate with one another to achieve their goals. The term "orchestrator" implies a hierarchical structure among agents, which contradicts our design philosophy. Thus, we named this special type of agent the "Ensemble Agent."
}

\small
\begin{minted}{yaml}
<ensemble_agent_name>: 
  type: <string>
  args: <array>
  contains: 
    - <agent_name>:
      args:
        - <arg_name in agent>: <arg_name in ensemble agent>
    - <agent_name>
  
  steps(optional):
    - <step>
\end{minted}
\normalsize

An ensemble agent can execute statements defined in the \textbf{steps} list at the beginning of a conversation and then run agents in the \textbf{contains} list.   When invoked, the ensemble agent analyzes the entire dialogue and, based on the current context, recommends the next action-such as selecting an appropriate agent, clarifying user input, providing a fallback response, terminating the conversation, or waiting for the user's reply.

% \begin{minted}{yaml}
% ensemble_agent_name:
%     type: ensemble agent
%     description: <string>
%     prompt: <string>
%     contains: <array>
%     steps: <array>
%     fallback: <string> | <object>
%     exit: <string> | <object>
% \end{minted}

The \textbf{contains} keyword lists all the agents managed and coordinated by the ensemble agent. The \textbf{args} field represents arguments specific to the ensemble agent. Synchronizing values between arguments in the ensemble agent and the arguments in agents managed by the ensemble agent can be specified as follows:

\small
\begin{minted}{yaml}
    - <arg_name in agent>: <arg_name in ensemble agent>
\end{minted}
\normalsize

If we allow the agent to change the value of the arguments in ensemble, we could put keyword \textbf{ref} inside,
\small
\begin{minted}[escapeinside=@@]{yaml}
    - <arg_name in agent>: @\textcolor{keycolor}{\textbf{ref}}@ <arg_name in ensemble agent>
\end{minted}
\normalsize

\subsection{Agent ``main''}
Similar to programming languages, ADL requires an explicit entry point. We define a special keyword, \textbf{main}, representing a unique agent that declares the chatbot's initialization process. It can be any type of agent named \texttt{main}. \nop{The \textbf{steps} field will initialize and then  invoke previously defined agents. This declaration uses the \textbf{call} field, allowing it to reference any previously defined agent.}

\small
\begin{minted}{yaml}
main: 
  type: flow agent | ensemble agent | llm agent | kb agent
  ...
\end{minted}
\normalsize

\section{Interfacing with External Services}
ADL supports interacting with external services. This can be achieved in multiple ways: (1) \textbf{Custom Function} is commonly used when retrieving information from sources such as databases, APIs, or when updating internal states. These functions are called within the \textbf{steps} statements. 
(2) \textbf{Tool Calling} allows the integration of custom functions within an LLM agent. Since an LLM agent may request tool execution during a conversation, ADL provides a mapping mechanism that minimizes the effort required for declaration.
(3) \textbf{Third-Party Agents}, through a standardized interface, can take over the conversation when necessary, enabling seamless integration with external systems.

\subsection{Custom Function}
For all functions mentioned in the agent declaration, there must be a corresponding Python implementation.  ADL uses the keyword \textbf{tools}, followed by a list of filenames, to import all relevant Python  scripts. 

\small
\begin{minted}{yaml}
tools:
  - <script_file_name>
\end{minted}
\normalsize
\nop{
\begin{minted}{yaml}
call: <function_name>
args: <array> 
\end{minted}
}

A custom function has to follow a specific format so that the caller can interpret the arguments to the function and the return value of the function correctly. 

\small
\begin{minted}{python}
def custom_function(required_arg: str, optional_arg: int = 0):
    """The description of this custom function"""
    ...
    print("message to the caller agent.")
    return [{"status": "success" | "error", "msg": "execution status message."},
            {"bot": "a response to the user."}, 
            {"arg": "argument_name", "value": 100}]
\end{minted}
\normalsize

The function arguments are defined explicitly, with optional parameters given default values. The return value follows a structured format: The first item is the last message sent to the caller agent.  The second item contains the bot's response text, and the third item updates arguments, specifying their name and assigned value.

Regarding the function output, the standard print output is sent to the caller agent. The function will have its official return that contains the bot response to the user.  This distinction ensures that function/tool execution results remain separate from bot responses.

\nop{
Furthermore, the return value mechanism allows functions to directly output any sentence during execution and return any updated arguments. To ensure consistency in function implementation and output formatting, follow the example below when defining custom functions:
}

\subsection{Tool Calling}
Tool calling occurs when an LLM  agent needs to invoke an external function. In this scenario, the function name should be explicitly mentioned in the LLM agent's prompt. The function implementation follows the same format as custom function.

ADL reads the Python code and translates it into the format required by the LLM, eliminating the need of explicitly writing a function request that adheres to strict specifications. All parameters can follow Python's annotation conventions to define their types. They are parsed by ADL and passed to the LLM as attributes. The function's docstring serves as its description and is also parsed by ADL and transmitted to the LLM.

\nop{When defining function arguments: 1) Explicitly defined parameters are required when the LLM calls the function. 2)Parameters with default values are optional and do not need to be specified.
} 

\subsection{Third-party Agents}
ADL can be extended in the future to support third-part agents.  By leveraging custom  functions, ADL can directly supports integration with third-party agents that do not interact with users during the execution.  For interactive agents, it can be achieved by adding another agent type, external agent, into the existing four agent types. This will enable sharing the conversation context and state with the external agents in a manner similar to internal agents. 
\nop{
In principle, ADL can interoperate with any agent framework, enabling seamless collaboration between different agent-based systems.
}

\section{Retail Banking Bot}
\label{banking}
We implement the same retail banking bot with ADL and Swarm. Here's the implementation detail. The code related to the database was not included.
\subsection{ADL Implementation}

\small
\begin{minted}[breaklines=true, breakanywhere=true, escapeinside=@@]{yaml}
transfer_money:
  type: llm agent
  description: Guides the user through the process of initiating a bank transfer.
  prompt: |  
    1. First, confirm that you understand the user wants to transfer money.
    2. Ask for the username. Based on the username, call the function "action_ask_account_from" to display the account name and balance. Ask which account to use and fill in account_from with the corresponding number (do not show the account number to the user, just populate account_from directly).
    3. Ask for payee_name.
    4. Prompt the user to enter the "add payee" agent and exit this agent.
    5. Once all required details are collected, call "action_check_sufficient_funds".
    6a. If there are sufficient funds, proceed to step 7.
    6b. If there are insufficient funds, terminate this agent immediately.
    7. Collect the transfer timing timing.
    8a. If timing is "now", confirm whether to proceed with the immediate transfer.
     - If confirmed, call "action_process_immediate_payment", then output "Transfer successful", and exit this agent.
     - If not proceeding immediately, output "Transfer canceled", and exit this agent.
    8b. If timing is not "now", ask for the specific transfer date, ensuring it is formatted as "DD/MM/YYYY". Proceed to step 9.
    9. Call "action_validate_payment_date".
     - If the date is a future date, confirm whether to schedule the transfer.
     - If confirmed, call "action_schedule_payment", output "Transfer scheduled", and exit this agent.
     - If not scheduling, output "Transfer canceled", and exit this agent.
  args:
    - username
    - account_from
    - payee_name
    - amount
    - timing
    - payment_date
  uses:
    - action_check_payee_existence
    - action_check_sufficient_funds
    - action_process_immediate_payment
    - action_validate_payment_date
    - action_schedule_payment

kb:
  type: kb agent
  file: path/to/docs

what_can_you_do:
  type: flow agent
  description: This flow addresses user inquiries about the assistant's capabilities.
  steps:
    - bot: "I am your Banking assistant. I can help you with transferring money, managing authorised payees, checking an account balance, blocking a card, and answering your general finance enquiries."

remove_payee:
  type: llm agent
  description: delete an existing authorised payee
  prompt: |
    This flow guides the user through specifying the payee to be removed, 
    call "action_remove_payee", and provides appropriate feedback on the success or failure of the operation.
  args:
    - payee_name
    - username
  uses:
    - action_remove_payee

list_payees:
  type: llm agent
  description: "Retrieve and display the user's list of authorized payees, allowing them to view all accounts available for transactions."
  prompt: |
    After knowing "username", call function "action_list_payees", Retrieve and display the user's list of authorized payees. The task ends.
  uses:
    - action_list_payees
  args:
    - username

check_balance:
  type: llm agent
  description: check an account balance
  prompt: |
    1. You need to ask for the username. Based on the username, call the function "action_ask_account" to display the account information to the user and ask which account name they want to inquire about.
    2. Based on the user's response, fill the variable "account" with the corresponding account_number, then call "action_check_balance" to inform the user of the balance for the requested account.
  args:
    - username
    - account
  uses:
    - action_ask_account
    - action_check_balance

block_card:
  type: flow agent
  description: Block or freeze a user's debit or credit card
  args:
    - reason_for_blocking
    - physical_address
    - fraud_reported
    - temp_block_card
  steps:
    - bot: "Okay, we can block a card. Let's do it in a few steps"
    - bot: "Please tell us the reason for blocking"
    - @\textcolor{keycolor}{\textbf{user}}@
    - if: the user claims "My card is damaged", "My card has expired"
      then: 
        - bot: "Thank you for letting us know. I'm sorry to hear the card was 
        ${reason_for_blocking}"
        - next: confirm_issue_new_card
    - else if: the user claims "I lost my card", "I suspect fraud on my account"
      then:
        - bot: "As your card was potentially stolen, it's crucial to report this
        incident to the authorities. Please contact your local law enforcement 
        agency immediately."
        - bot: "Since you have reported ${reason_for_blocking}, we will block 
        your card"
        - next: confirm_issue_new_card
    - else if: the user claims "I'm planning to travel soon", "I'm moving to a 
      new address"
      then:
        - bot: "Thanks for informing us about moving."
        - bot: "Since you are ${reason_for_blocking}, we will temporarily block
        your card."
        - next: confirm_issue_new_card
      else:
        - bot: "Should you require further assistance, please contact our support
        team at 020 7777 7777. Thank you for being a valued customer."
        - call: action_update_card_status  
        
  confirm_issue_new_card:
    - bot: "Would you like to be issued a new card?"
    - @\textcolor{keycolor}{\textbf{user}}@
    - if: the user claims "Yes, send me a new card"
      then:
        - next: retrieve_user_address
      else:
        - call: action_update_card_status
        
  retrieve_user_address:
    - bot: "I have found your address: ${physical_address}. Should the new card 
    be delivered there?"
    - @\textcolor{keycolor}{\textbf{user}}@
    - if: the user claims "Yes"
      then:
        - bot: "Your card will be delivered to ${physical_address} within 7 
        business days"
      else:
        - bot: "Should you require further assistance, please contact our support 
        team at 020 7777 7777. Thank you for being a valued customer."
    - call: action_update_card_status  


add_payee:
  type: llm agent
  description: Add a new payee to the user's account
  prompt: |
    1. Prompt the user sequentially for payee_name, account_number, payee_type (person/business), and reference (a short note to identify the payee or purpose).
    2. Confirm whether all the provided information is correct. If correct, call "action_add_payee".
    3. If the action is successful, respond with "{payee_name} has been successfully added to your list of authorized payees." Otherwise, respond with "I'm sorry, but there was an error adding {payee_name}. Please try again later or contact Customer Support.
    4. Ask the user if they need to make a transfer; if so, enter the transfer_money agent.
  args:
    - username
    - payee_name
    - account_number
    - payee_type
    - reference
  uses:
    - action_add_payee

meta:
  type: ensemble agent
  contains:
    - what_can_you_do
    - check_balance:
        args:
          username: ref username
    - remove_payee:
        args:
          username: ref username
    - transfer_money:
        args:
          username: ref username
    - list_payees:
        args:
          username: ref username
    - block_card:
        args:
          username: ref username
          physical_address: ref physical_address
    - add_payee:
        args:
          username: ref username
  args:
    - username
    - segment
    - email_address
    - physical_address
  steps:
    - set:
        username: "Mary Brown"
    - call: action_session_start
      args:
        username: username
    - set:
        segment: action_session_start.segment
        email_address: action_session_start.email_address
        physical_address: action_session_start.physical_address

main:
  type: flow agent
  steps:
    - call: meta

\end{minted}
\normalsize

\subsection{Swarm Implementation}

\small
\begin{minted}[breaklines=true, breakanywhere=true]{python}
def transfer_to_greeting():
    return greeting

def transfer_to_add_payee():
    return add_payee

def transfer_to_check_balance():
    return check_balance

def transfer_to_list_payees():
    return list_payees

def transfer_to_remove_payee():
    return remove_payee

def transfer_to_transfer_money():
    return transfer_money

def transfer_to_get_feedback():
    return get_feedback

def transfer_to_block_card():
    block_card()


def transfer_to_triage():
    """Call this function when a user needs to be transferred to a different agent and a different policy.
    For instance, if a user is asking about a topic that is not handled by the current agent, call this function.
    """
    return triage_agent

def block_card(context_variables):
    def confirm_issue_new_card():
        print("Would you like to be issued a new card?")
        user = input().strip()
        if "Yes" in user:
            retrieve_user_address()
        else:
            action_update_card_status()

    def retrieve_user_address():
        print(f"I have found your address: {context_variables.get('physical_address')}. Should the new card
be delivered there?")
        user = input().strip()
        if "Yes" in user:
            print(f"Your card will be delivered to {context_variables.get('physical_address')} within 7 business days")
        else:
            print("Should you require further assistance, please contact our support team at 020 7777 7777. Thank you for being a valued customer.")
        action_update_card_status()
    
    print("Okay, we can block a card. Let's do it in a few steps")
    print("Please tell us the reason for blocking")

    condition = Agent(
        name="blocking_condition",
        instructions="Your task is to identify the user's reason for blocking a card. The possible reasons are: lost, damaged, stolen, suspected of fraud, malfunctioning, or expired. Based on the user input, return the corresponding reason. If none of the above apply, return 'other'."
    )

    client = Swarm()
    
    while True:
        user = input().strip()
        messages = [{"role": "user", "content": user}]
        
        if "exit" in user or "quit" in user:
            break
        
        reason = client.run(condition, messages)
        if "damaged" in reason or "expired" in reason:
            print(f"Thank you for letting us know. I'm sorry to hear the card was {reason}")
            confirm_issue_new_card()
        elif "lost" in reason or "suspected of fraud" in reason:
            print("As your card was potentially stolen, it's crucial to report this incident to the authorities. Please contact your local law enforcement agency immediately.")
            print("Since you have reported {reason}, we will block your card")
            confirm_issue_new_card()
        elif "malfunctioning" in reason:
            print("Thanks for informing us about moving.")
            print("Since you are {reason}, we will temporarily block your card.")
            confirm_issue_new_card()
        else:
            print("Should you require further assistance, please contact our support team at 020 7777 7777. Thank you for being a valued customer.")
            action_update_card_status()
        
def triage_instructions(context_variables):
    customer_context = context_variables.get("customer_context", None)

    return f"""You are to triage a users request, and call a tool to transfer to the right intent.
    Once you are ready to transfer to the right intent, call the tool to transfer to the right intent.
    You dont need to know specifics, just the topic of the request.
    When you need more information to triage the request to an agent, ask a direct question without explaining why you're asking it.
    Do not share your thought process with the user! Do not make unreasonable assumptions on behalf of user.
    The customer context is here: {customer_context}"""
    
triage_agent = Agent(
    name="Triage Agent",
    instructions=triage_instructions,
    functions=[
               transfer_to_greeting,
               transfer_to_add_payee,
               transfer_to_block_card,
               transfer_to_check_balance,
               transfer_to_list_payees,
               transfer_to_remove_payee,
               transfer_to_transfer_money,
               ],
)


GREETING_PROMPT = """Answer: I am your Banking assistant. I can help you with transferring money, managing authorised payees, checking an account balance, blocking a card, and answering your general finance enquiries."""

ADD_PAYEE_PROMPT = """1.	Prompt the user sequentially for payee_name, account_number, payee_type (person/business), and reference (a short note to identify the payee or purpose).
    2.	Confirm whether all the provided information is correct. If correct, call "action_add_payee".
    3.	If the action is successful, respond with "{payee_name} has been successfully added to your list of authorized payees." Otherwise, respond with "I'm sorry, but there was an error adding {payee_name}. Please try again later or contact Customer Support.
    4. Ask the user if they need to make a transfer; if so, enter the transfer_money agent."
"""

GET_FEEDBACK_POLICY = """Ask the user if they are satisfied with the service. voting from 1 to 5. If it is less than 3, set satisfied = True; otherwise, set satisfied = False"""

CHECK_BALANCE_POLICY = """1.	You need to ask for the username. Based on the username, call the function "action_ask_account" to display the account information to the user and ask which account name they want to inquire about.
    2.	Based on the user's response, fill the variable "account" with the corresponding account_number, then call "action_check_balance" to inform the user of the balance for the requested account."""

LIST_PAYEES_POLICY = """After knowing "username", call function "action_list_payees", Retrieve the user's list of authorized payees. Call "get_feedback" to get the feedback.The task ends."""



TRANSFER_MONEY_POLICY = """
0. Check satisfied value. If satisfied == False, express your apology and jump to step 2. Otherwise, proceed to step 1.
1.	First, confirm that you understand the user wants to transfer money.
    2.	Ask for the username. Based on the username, call the function "action_ask_account_from" to display the account name and balance. Ask which account to use and fill in account_from with the corresponding number (do not show the account number to the user, just populate account_from directly).
    3.	Ask for payee_name.
    4. Prompt the user to enter the "add payee" agent and exit this agent.
    5.	Once all required details are collected, call "action_check_sufficient_funds".
    6a. If there are sufficient funds, proceed to step 7.
    6b. If there are insufficient funds, terminate this agent immediately.
    7.	Collect the transfer timing timing.
    8a. If timing is "now", confirm whether to proceed with the immediate transfer.
    8b. If timing is not "now", ask for the specific transfer date, ensuring it is formatted as "DD/MM/YYYY". Proceed to step 9.
    9.	Call "action_validate_payment_date".
    9a. If the date is a future date, confirm whether to schedule the transfer.
    9b. If confirmed, call "action_schedule_payment", output "Transfer scheduled", and exit this agent.
    9c. If not scheduling, output "Transfer canceled", and exit this agent."""

REMOVE_PAYEE_POLICY = """This flow guides the user through specifying the payee to be removed.
        1. call "list_payees_agent" to get the current_payee_number. Set it as previous_payee_number.
        2. call "action_remove_payee".
        3. call "list_payees_agent" again to get the current_payee_number.
        4. if current_payee_number - previous_payee_number == 0, tell that"Failed to removed!"; otherwise, tell that"Succeed removed!"""

greeting = Agent(
    name="greeting",
    instructions=GREETING_PROMPT,
    functions=[transfer_to_triage]
)

add_payee = Agent(
    name="Add Payee",
    instructions=ADD_PAYEE_PROMPT,
    functions=[action_add_payee, transfer_to_triage],
)

check_balance = Agent(
    name="Check Balance",
    instructions=CHECK_BALANCE_POLICY,
    functions=[actio_ask_account, action_check_balance, transfer_to_triage],
)

list_payees = Agent(
    name="List Payees",
    instructions=LIST_PAYEES_POLICY,
    functions=[action_list_payees, transfer_to_triage, transfer_to_get_feedback],
)

remove_payee = Agent(
    name="Remove Payee",
    instructions=REMOVE_PAYEE_POLICY,
    functions=[action_remove_payee, transfer_to_triage],
)

transfer_money = Agent(
    name="Transfer Money",
    instructions=TRANSFER_MONEY_POLICY,
    functions=[action_check_payee_existence, action_check_sufficient_funds, action_process_immediate_payment, action_schedule_payment, action_validate_payment_date, transfer_to_triage],
)


\end{minted}
\normalsize

\subsection{Static Debugging Prompt}
\label{prompt}

\begin{table*}[htbp]
\centering
\begin{tabularx}{\textwidth}{X}

\hline

<ADL or Swarm code here>

Is there any infinite loop in the logic of this chatbot? If so, please point it out. Otherwise, simply reply "No." \\

\hline
\end{tabularx}
\caption{Debugging Prompt}
\label{tab:debug-prompt}
\end{table*}

\newpage

\section{A Book Store Chatbot}\label{app3}
This example covers nearly all the syntactic features of ADL.  It is a bookstore chatbot supporting return policy inquiries, online ordering, and product recommendations. 

The \texttt{store\_policy\_kb} agent aggregates information from internal documentation, website content, and frequently asked questions. If a user's issue can be addressed in any of these sources, the \texttt{store\_policy\_kb} agent retrieves the corresponding answer.

The \texttt{book\_recommendation} agent interacts directly with users, inquiring about their book preferences, querying the product database based on the specified genres, and presenting recommendations selected from top-selling books.

Finally, the \texttt{order} agent manages the purchase process by confirming all items the user intends to buy and executing the order. If the user wishes to select additional books, the agent invokes the \texttt{book\_recommendation} agent accordingly.

\small
\begin{minipage}[t]{0.5\linewidth}
\begin{minted}[breaklines]{yaml}
triage:
  type: ensemble agent
  description: Select an agent to respond to the user.
  args:
    - date
  contains:
    - store_policy_kb
    - book_recommendation
    - order
  steps:
    - call: get_date
    - set:
        date: get_date.today
    - bot: "Hi, I'm your bookstore 
    assistant. How can I help you?"
  exit: default
\end{minted}
\end{minipage}
\hfill
\begin{minipage}[t]{0.5\linewidth}
\begin{minted}[breaklines]{yaml} 
main:
  type: flow agent
  steps:
    - call: triage
    
order:
  type: flow agent
  description: I can place an order.
  args:
    - books
    - order_status
  fallback: "Provide a fallback message: \"Sorry, I didn't understand that. Could you rephrase it?\" If the fallback occurs three times consecutively, terminate the conversation."
\end{minted}
\end{minipage}
\normalsize

\small
\begin{minipage}[t]{0.48\linewidth}
\begin{minted}[escapeinside=@@, breaklines]{yaml}
store_policy_kb:
  type: kb agent
  description: I can answer questions related to the store's policy.
  sources:
    - return_policy.pdf
    - www.bookstore.com/help-center
  faq:
    - q: Thanks, bye
      a: Looking forward to serving you next time.  
book_recommendation:
  type: llm agent
  description: I can recommend books to customers.
  args:
    - genre
    - book_info
  prompt: |
    1. Ask the user if they have a 
    preferred book genre.
    2. If the user has a favorite 
    book, call the "query_book_genre"
    function based on their 
    response to obtain the "genre".
    3. Using the genre and other 
    attributes, call the "find_best
    sellers" function to recommend 
    relevant books to the user.
  uses:
    - query_book_genre
    - find_bestsellers
\end{minted}
\end{minipage}
\hfill
\begin{minipage}[t]{0.48\linewidth}
\begin{minted}[escapeinside=@@, breaklines]{yaml}
  steps:
    - bot: "I'll place the order for you."
    - label: confirm_books
    - bot: "You have selected these books so far: ${books}. Would you like to add anything else to your order?"
    - @\textcolor{keycolor}{\textbf{user}}@
    - if: the user claims "Do you have other types of books?"
      then:
        - call: book_recommendation
        - next: confirm_books
    - else if: the user claims "I don't have anything else I want to buy."
      then:
        - next: start_ordering_operation
      else:
        - call: store_policy_kb
        - next: confirm_books
          tries: 3
        
  start_ordering_operation:
    - call: place_order
      args:
        ordered_book: books
        date: triage.date
    - if: place_order.status == True
      then:
        - return: success, Order placed successfully.
      else:
        - return: error, Order failed.
        

\end{minted}
\end{minipage}
\normalsize

\section{Case Study}
\label{rasa}
Rasa \citep{bocklisch2017rasa} is an open source machine learning framework to automate text and voice-based conversations. Its open source version (\url{https://github.com/RasaHQ/rasa}) is well known and popularly used.  Recently, it adopted the CALM architecture \citep{bocklisch2024} to integrate large language models into its  pipeline. Rasa-CALM follows the same design philosophy, explicitly requiring clear definitions for all components, including dialogue flows, bot responses, and conversation states. It strictly follows the predefined logic to reduce hallucination. At the same time, by incorporating LLMs, it gains a stable text comprehension ability.  \nop{ In theory, if users interact with the chatbot strictly according to the predefined flow, Rasa-CALM can execute successfully.}

Below is an example from Rasa, which showcases a \textit{Retail Banking} bot (\url{https://github.com/rasa-customers/starterpack-retail-banking-en}). It includes multiple functionalities such as transferring money, checking balances, managing payees, and blocking cards. In the following section, we will show the implementation of these functionalities in ADL and Rasa-CALM.

\subsection{Remove payee}

The process of removing a payee involves prompting the user for the payee's name, attempting a removal operation in the database, and then responding to the user based on the outcome. CALM divides chatbot definitions into four distinct parts: business logic (\textbf{flows}), bot responses (\textbf{responses}), external services (\textbf{actions}), and the bot's internal states (\textbf{slots}). 

For the remove payee functionality, CALM first defines the dialogue logic under the \textbf{flows} field, specifying required arguments, bot responses, and interaction timing with external services. Next, it defines an indicator slot in the \textbf{slots} field to serve as a flag returned by the executed action. Additionally, all bot responses involved in the conversation are explicitly listed under the \textbf{responses} keyword. Finally, the \textbf{actions} field enumerates the external functions needed for the procedure-in this example, only \texttt{action\_remove\_payee} is required.
Below is the implementation of this procedure using Rasa-CALM:

\small
\begin{minted}[breaklines]{yaml}
flows:
  remove_payee:
    if: False
    description: |
      Facilitates the process of removing an existing payee from a user's
      account, ensuring the user is authenticated before proceeding. This flow
      guides the user through specifying the payee to be removed, attempts the
      removal action, and provides appropriate feedback on the success or
      failure of the operation
    name: delete an existing authorised payee
    steps:
      - collect: payee_name
      - action: action_remove_payee
        next:
          - if: slots.payee_removed
            then:
              - action: utter_payee_removed_success
                next: END
          - else:
              - action: utter_payee_removed_failure
                next: END
slots:
  payee_removed:
    type: bool

responses:
  utter_ask_payee_name_to_remove:
    - text: "Which payee would you like to remove?"
  utter_payee_removed_success:
    - text: "{payee_name} has been successfully removed from your list of authorised payees"
  utter_payee_removed_failure:
    - text: "I'm terribly sorry, but there was an error removing {payee_name}. Please try again later or contact Customer Support"
actions:
  - action_remove_payee
\end{minted}
\normalsize

Correspondingly, ADL regards these components as various properties of an independent agent. Specifically, ADL describes the dialogue logic in natural language, placing it within the \textbf{prompt} field. The conversation-related arguments (\texttt{payee\_name} and \texttt{username}, in this example) are explicitly defined within the \textbf{args} field, and external functions required are listed under the \textbf{uses} keyword. Below is the equivalent definition of the remove-payee functionality in ADL:

\small
\begin{minted}[breaklines]{yaml}
remove_payee:
  type: llm agent
  description: delete an existing authorised payee
  args:
    - payee_name
    - username
  prompt: |
    This flow guides the user through specifying the payee to be removed, call 
    "action_remove_payee", and provides appropriate feedback on the success or 
    failure of the operation.
  uses:
    - action_remove_payee
\end{minted}
\normalsize

A key feature of ADL is that it lets you describe business logic in natural language. If needed, you can also define the same logic using a flow agent in ADL. The example below shows how the logic can be written as a list of steps:

\small
\begin{minted}[escapeinside=@@, breaklines]{yaml}
remove_payee:
  type: flow agent
  description: "Agent responsible for removing a payee from user's account."
  args:
    - payee_name
    - username
  steps:
    - bot: "Please provide the name of the payee you wish to remove."
    - @\textcolor{keycolor}{\textbf{user}}@
    - call: action_remove_payee
      args:
        payee_name: payee_name
        username: username
    - if: action_remove_payee.status == 'success'
      then:
        - bot: "${payee_name} has been successfully removed from your list of 
        authorised payees"
      else:
        - bot: "I'm terribly sorry, but there was an error removing ${payee_
        name}. Please try again later or contact Customer Support"
\end{minted}
\normalsize

\subsection{Card blocking}

Card blocking is another functionality that involves multiple arguments and a more complex workflow. In the CALM implementation, the conversation logic is defined under the \textbf{flows} field as follows: When a user requests card blocking, the bot first asks for the reason and then adopts different response strategies based on the provided reason.

If the user indicates that her card is damaged or expired, the bot acknowledges this, confirms whether to issue a new card, and marks the old card as inactive by calling the function \texttt{action\_update\_card\_status()}. If the card is lost or stolen, the bot first blocks the old card, proceeds to the new card issuance process, and then marks the old card as inactive. If the reason is traveling or moving, the bot temporarily blocks the old card and proceeds to issue a new card. For other reasons, the bot directly instructs the user to contact support by phone.

\subsubsection{CALM implementation}

The following is an implementation using CALM:

\small
\begin{minted}[breaklines]{yaml}
flows:
  block_card:
    description: "Block or freeze a user's debit or credit card to prevent unauthorized use, stop transactions, or report it lost, stolen, damaged, or misplaced 
    for added security"
    name: block a card
    steps:
      - action: utter_block_card_understand
      - call: select_card
      - collect: reason_for_blocking
        description: |
          The reason for freezing or blocking the card, described as lost, damaged, stolen, suspected of fraud, malfunctioning, or expired. The user may say they are traveling or moving, or they may say they want to temporarily freeze their card. For all other responses, set reason_for_blocking slot to 'unknown'.
        next:
          - if: "slots.reason_for_blocking == 'damaged' or slots.reason_for_blocking == 'expired'"
            then: "acknowledge_reason_damaged_expired"
          - if: "slots.reason_for_blocking == 'fraud' or slots.reason_for_blocking == 'stolen' or slots.reason_for_blocking == 'lost'"
            then:
              - set_slots:
                  - fraud_reported: true
                next: "acknowledge_reason_fraud_stolen_lost"
          - if: "slots.reason_for_blocking == 'traveling' or slots.reason_for_blocking == 'moving'"
            then:
              - set_slots:
                  - temp_block_card: true
                next: "acknowledge_reason_travelling_moving"
          - else: "contact_support"
      - id: acknowledge_reason_damaged_expired
        action: utter_acknowledge_reason_damaged_expired
        next: "confirm_issue_new_card"
      - id: acknowledge_reason_fraud_stolen_lost
        action: utter_acknowledge_reason_fraud_stolen_lost
        next: "card_blocked"
      - id: acknowledge_reason_travelling_moving
        action: utter_acknowledge_reason_travelling_moving
        next: "card_blocked"
      - id: "card_blocked"
        action: "utter_card_blocked"
        next: "confirm_issue_new_card"
      - id: "confirm_issue_new_card"
        collect: confirm_issue_new_card
        description: |
          Confirm if the user wants to be issued a new card. The answer should
          be an affirmative statement, such as "yes" or "correct," or a declined 
          statement, such as "no" or "I don't want to"
        ask_before_filling: true
        next:
          - if: "slots.confirm_issue_new_card"
            then: "retrieve_user_address"
          - else: "update_card_status"
      - id: "retrieve_user_address"
        collect: address_confirmed
        description: |
          Confirm if the given address is correct. The answer should be an affir-
          mative statement, such as "yes" or "correct," or a declined statement, 
          such as "no" or "that's not right."
        next:
          - if: "slots.address_confirmed"
            then: "card_sent"
          - else: "contact_support"
      - id: "card_sent"
        action: utter_confirm_physical_address
        next: update_card_status
      - id: "contact_support"
        action: utter_contact_support
        next: update_card_status
      - id: "update_card_status"
        action: action_update_card_status
        next: END

slots:
  reason_for_blocking:
    type: categorical
    values:
      - lost
      - fraud
      - stolen
      - damaged
      - expired
      - traveling
      - moving
  address_confirmed:
    type: bool
  fraud_reported:
    type: bool
    initial_value: false
  temp_block_card:
    type: bool
    initial_value: false
  confirm_issue_new_card:
    type: bool
  address:
    type: text
  card_status:
    type: categorical
    values:
      - active
      - inactive
actions:
  - action_update_card_status

responses:
  utter_ask_reason_for_blocking:
    - text: "Please tell us the reason for blocking"
      buttons:
      - title: "I lost my card"
        payload: "/SetSlots(reason_for_blocking=lost)"
      - title: "My card is damaged"
        payload: "/SetSlots(reason_for_blocking=damaged)"
      - title: "I suspect fraud on my account"
        payload: "/SetSlots(reason_for_blocking=fraud)"
      - title: "My card has expired"
        payload: "/SetSlots(reason_for_blocking=expired)"
      - title: "I'm planning to travel soon"
        payload: "/SetSlots(reason_for_blocking=traveling)"
      - title: "I'm moving to a new address"
        payload: "/SetSlots(reason_for_blocking=moving)"
  utter_block_card_understand:
    - text: "Okay, we can block a card. Let's do it in a few steps"
      metadata:
        rephrase: True
  utter_ask_address_confirmed:
    - text: "I have found your address: {physical_address}. Should the new card
    be delivered there?"
      buttons:
        - title: "Yes"
          payload: "/SetSlots(address_confirmed=True)"
        - title: "No"
          payload: "/SetSlots(address_confirmed=False)"
  utter_confirm_physical_address:
    - text: "Your card will be delivered to {physical_address} within 7 business days"
  utter_card_blocked:
    - condition:
        - type: slot
          name: fraud_reported
          value: true
      text: "Since you have reported {reason_for_blocking}, we will block your card"
    - condition:
        - type: slot
          name: temp_block_card
          value: true
      text: "Since you are {reason_for_blocking}, we will temporarily block your 
      card."
    - text: We will block your card.
  utter_ask_confirm_issue_new_card:
    - text: "Would you like to be issued a new card?"
      buttons:
        - title: "Yes, send me a new card"
          payload: "/SetSlots(confirm_issue_new_card=true)"
        - title: "No, just block my card"
          payload: "/SetSlots(confirm_issue_new_card=false)"
  utter_ask_address:
    - text: "Would you like us to deliver your new card to this address: 
    {physical_address}?"
      buttons:
        - title: "Yes, send a new card"
          payload: "/SetSlots(address_confirmed=true)"
        - title: "No, I'll go to the bank"
          payload: "/SetSlots(address_confirmed=false)"
  utter_contact_support:
    - text: "Should you require further assistance, please contact our support team at 020 7777 7777. Thank you for being a valued customer."
    - text: "If you have any questions or concerns, please don't hesitate to reach out to our support team at 020 7777 7777. We're here to help."
    - text: "For additional support, please contact our customer service team at 020 7777 7777. Thank you for being a valued customer."
  utter_acknowledge_reason_damaged_expired:
    - text: "Thank you for letting us know. I'm sorry to hear the card was {reason_for_blocking}"
      metadata:
        rephrase: True
  utter_acknowledge_reason_fraud_stolen_lost:
    - text: "As your card was potentially stolen, it's crucial to report this incident to the authorities. Please contact your local law enforcement agency immediately."
    - text: "Given the unfortunate potential theft of your card, please report this incident to your local law enforcement agency. We'll work together to minimize the impact of this situation."
  utter_acknowledge_reason_travelling_moving:
    - text: Thanks for informing us about moving.
\end{minted}
\normalsize

\subsubsection{ADL implementation}
Similarly, ADL can implement the same functionality using either an LLM agent or a Flow agent. Below is the LLM agent version, where all conversation logic and bot responses are described into natural language within the \textbf{prompt} field:

\small
\begin{minted}[breaklines]{yaml}
block_card:
  type: llm agent
  description: ask for reason and block the card
  args:
    - username
    - card
    - physical_address
    - reason_for_blocking
  prompt: |
    You are an agent assisting users in blocking their cards.
    1. Based on the username, directly call "action_ask_card" to retrieve the 
    user's available cards. Ask the user: "Select the card you require assistance 
    with:" and fill in the "card" field.  
    2. Ask the user for the reason they want to block the card (reason_for_blocking):  
       - The reason should be described as lost, damaged, stolen, suspected of 
       fraud, malfunctioning, or expired.  
       - The user may also say they are traveling or moving, or that they want 
       to temporarily freeze their card.  
       - For any other responses, set the "reason_for_blocking" slot to "un-
       known".  
    3a. If "reason_for_blocking" is "damaged" or "expired", inform the user:
       "Thank you for letting us know. I'm sorry to hear the card was ${reason_
       for_blocking}." Proceed to Step 4.  
    3b. If "reason_for_blocking" is "fraud", "stolen", or "lost", inform the 
        user:  "As your card was potentially stolen, it's crucial to report this
        incident to the authorities. Please contact your local law enforcement 
        agency immediately."  Proceed to Step 4.  
    3c. If "reason_for_blocking" is "traveling" or "moving", inform the user:  
       "Thanks for informing us about moving."  Proceed to Step 4.  
    3d. For all other cases, instruct the user to contact "020 7777 7777" and 
    call the function "action_update_card_status", then end the agent process.  
    4. Inform the user that their card will be blocked due to the specified reason.  
       - Confirm whether they want to **issue a new card** or if they prefer to 
       visit the bank themselves.  
       - If the user confirms they want to issue a new card, proceed to Step 5.  
       - If the user declines, call "action_update_card_status" and end the agent 
       process.  
    5. Confirm the user's "physical_address".  
       - If the user confirms the address is correct, inform them:  
         "The new card will be sent to this address."  
       - If the user indicates the address is incorrect, go back to Step 3d.  
  uses:
    - action_ask_card
    - action_update_card_status
\end{minted}
\normalsize

The Flow agent implementation provides a clear, step-by-step illustration of the same procedure, explicitly defining the logic flow:

\small
\begin{minted}[escapeinside=@@,breaklines]{yaml}
block_card_flow:
  type: flow agent
  description: Block or freeze a user's debit or credit card
  args:
    - reason_for_blocking
    - physical_address
    - fraud_reported
    - temp_block_card
  steps:
    - bot: "Okay, we can block a card. Let's do it in a few steps"
    - bot: "Please tell us the reason for blocking"
    - @\textcolor{keycolor}{\textbf{user}}@
    - if: the user claims "My card is damaged", "My card has expired"
      then: 
        - bot: "Thank you for letting us know. I'm sorry to hear the card was 
        ${reason_for_blocking}"
        - next: confirm_issue_new_card
    - else if: the user claims "I lost my card", "I suspect fraud on my account"
      then:
        - bot: "As your card was potentially stolen, it's crucial to report this
        incident to the authorities. Please contact your local law enforcement 
        agency immediately."
        - bot: "Since you have reported ${reason_for_blocking}, we will block 
        your card"
        - next: confirm_issue_new_card
    - else if: the user claims "I'm planning to travel soon", "I'm moving to a 
      new address"
      then:
        - bot: "Thanks for informing us about moving."
        - bot: "Since you are ${reason_for_blocking}, we will temporarily block
        your card."
        - next: confirm_issue_new_card
      else:
        - bot: "Should you require further assistance, please contact our support
        team at 020 7777 7777. Thank you for being a valued customer."
        - call: action_update_card_status  
        
  confirm_issue_new_card:
    - bot: "Would you like to be issued a new card?"
    - @\textcolor{keycolor}{\textbf{user}}@
    - if: the user claims "Yes, send me a new card"
      then:
        - next: retrieve_user_address
      else:
        - call: action_update_card_status
        
  retrieve_user_address:
    - bot: "I have found your address: ${physical_address}. Should the new card 
    be delivered there?"
    - @\textcolor{keycolor}{\textbf{user}}@
    - if: the user claims "Yes"
      then:
        - bot: "Your card will be delivered to ${physical_address} within 7 
        business days"
      else:
        - bot: "Should you require further assistance, please contact our support 
        team at 020 7777 7777. Thank you for being a valued customer."
    - call: action_update_card_status
\end{minted}
\normalsize

% The figture~\ref{fig:case} is a partial implementation of the payee removal procedure using ADL and Rasa-CALM. The left side shows the CALM implementation, which divides the entire chatbot definition into four distinct parts. This flow first collects the payee's name, then removes the corresponding entry from an external database, and finally selects an appropriate response from the return value of the function \textit{action\_remove\_payee}. 

% We implemented the same functionality using ADL on the right-hand side. In ADL, the implementation is integrated into a single agent using arguments and function calls. Additionally, ADL can realize this functionality using a flow declaration method similar to CALM.
\subsection{Conclusion}
Both methods have their own advantages and limitations. The state-declaration approach used by Rasa-CALM provides fine-grained control over dialogue content and progression. However, this comes at the cost of flexibility, making it difficult to handle unforeseen situations in real conversations.

In contrast, ADL simplifies the programming process by allowing logic to be expressed through natural language or intuitive flow structures. Whether this simplification results in a better implementation remains an open question.

An objective evaluation of the strengths and weaknesses of these implementation strategies is an ongoing challenge for researchers. A suitable benchmark for comparison is needed in addressing this question.

\end{document}